\documentclass[12pt]{article}

\usepackage{amsmath}
\usepackage{amssymb}
\usepackage{multicol}
\usepackage{tabularx}
\usepackage{euscript}
\usepackage{calc}
\usepackage{indentfirst}

\def\@PACS{}
\def\PACS#1{\gdef\@PACS{#1}}

\def\address#1{\expandafter\def\expandafter\@authoraddress\expandafter
  {\@authoraddress{\footnotesize\sl\centering\ignorespaces#1\unskip\par}}}

\begin{document}

\title{Energy spectrum of the equal spin-spin interactions Hamiltonian}

\author{A.R.~Kessel \thanks{kessel@dionis.kfti.knc.ru},
        R.R.~Nigmatullin, N.M.~Yakovleva}

%%% author's address(es)
\address{E.K.~Zavoisky Physical-Technical Institute RAS, 420029 Kazan, Russia}

\maketitle

\begin{abstract}
The energy spectrum of eigenvalues of the equal 
spin-spin interactions (ESSI) Hamiltonian has been found. 
The obtained spectrum is free from limitations imposed on 
number of spins and parameters of the ESSI Hamiltonian. 
This model can be used for consideration of spin dynamics 
of mesoscopic systems and molecules with large number 
of nuclei spins.
\end{abstract}

%%% PACS numbers

\section{Introduction. Many-body problem of spin-spin interactions}
Thorough attention of researches to the properties of the Hamiltonian 
of spin-spin interactions is kept during some decades. It is due to 
the fact that this Hamiltonian determines fine peculiarities of 
dynamics and kinetics of the electron and nuclei spins of paramagnet 
spin-systems. The Hamiltonian of a paramagnet (electronic and nuclear) 
has the following form  \cite{bib:1}
\begin{equation}
\mathcal{H} = \mathcal{H}_{z} + \mathcal{H}_{d} + \mathcal{H}_{dd},
\label{eqn:1}
\end{equation}
\begin{gather*}
\mathcal{H}_{z} = \hbar\omega_{0}\sum_{j}S^{z}_{j}, ~ \mathcal{H}_{d} = \hbar\sum_{fj}\mathrm{A}_{fj}S^{z}_{j}S^{z}_{f}, ~
\mathcal{H}_{dd} = \hbar\sum_{fj}\mathrm{B}_{fj}(S^{-}_{j}S^{+}_{f} + S^{+}_{j}S^{-}_{f}), \nonumber \\
S^{\pm} = S^{x}\pm iS^{y}, ~
\mathrm{A}_{fj} = A_{fj} + J_{fj}, ~ A_{fj} = (\hbar\gamma^{2}/2R_{fj}^{3})(1-3\cos\theta_{fj}), \nonumber \\
\mathrm{B}_{fj} = (-1/4)A_{fj} + (1/2) J_{fj},
\end{gather*}
where $S^{a}_{j}$ is the $a$-th spin component of a spin $S_{j}$ ($a = x, y, z$), 
located in the point $\mathbf{j}$, ($R_{fj}, \theta_{fj}, \varphi_{fj}$) define 
the spherical coordinates of a vector connecting the points $f$ and $j$ 
in the laboratory frame which $OZ$ axes is parallel to the applied 
constant magnetic field  $H_{0}$, $\gamma$ is a gyromagnetic ratio, 
$\omega_{0} = \gamma H_{0}$ is Zeeman frequency, the value  
$J_{fj}$ defines an exchange integral. Only the secular (diagonal) part 
of the dipole-dipole Hamiltonian is given by expression (\ref{eqn:1}). 
This part determines the most essential contribution to dynamics and 
kinetics of spins in comparison with off-diagonal parts \cite{bib:2,bib:3,bib:4}.
The Hamiltonian (\ref{eqn:1}) represents a typical example of many-body 
interactions and so it is no needless to say about exact 
diagonalization of (\ref{eqn:1}) if number of interacting spins $n \ge 10$.
Even if the numerical calculation of the corresponding eigenvalues 
will be obtained, the analysis of this information becomes difficult 
because the number of states $N =2^{n}$ is growing exponentially with 
the increasing of $n$. So in existing theories only macroscopic 
characteristics as spin-spin relaxation times, the second and 
the fourth moments of resonance lines were taken into 
account \cite{bib:1}.

Recently in investigation of the NMR on protons in a nematic 
liquid crystal \cite{bib:5} a principal new NMR spectrum has 
been demonstrated. It consists of a set of well-resolved 
resonance lines instead of one usual unresolved resonance 
line broadened by spin-spin interactions. In recent time 
the authors brought  the number of resonance lines up to 
the value 1024 \cite{bib:6}. This result opens new 
possibilities in studying of dynamics and kinetics 
of dipole-dipole connected spins in the mesoscopic 
systems and gives a new perspective physical medium 
for quantum information science.

In such a way, the theoretical description of quantum-statistical 
properties of dipole-dipole coupled spin-systems becomes a very 
interesting problem. 

A liquid crystal where the spin-spin multiplet resolved 
spectrum with narrow lines was observed  \cite{bib:5}, 
includes partly oriented molecules, each molecule 
contains $n=19$ interacting protons and intermolecular 
spin-spin interaction is negligible. It implies to find 
about a half-million of eigenvalues of the Hamiltonian (\ref{eqn:1})
that is the useless problem for exact solution. We want to 
stress that at certain values of orientation angles 
$\theta_{fj}$ the dipole-dipole interaction between 
close disposed spins can be less in comparison with 
remote spins. It is necessary to take also into account 
that orientation of spins in the space is not fixed 
precisely and true interaction constants are effectively 
replaced by their parts non-averaged by heat motion. 
This averaging of the spin-spin interaction of protons, 
located in the different part of the molecule is not 
equivalent: the effect of averaging is greater in 
the more mobile fragments. So, the further progress 
is possible with using a simplified theoretical model, 
which keeps nevertheless the main features of the 
basic Hamiltonian (\ref{eqn:1}). In our opinion such 
theoretical model can be presented by the 
\emph{equal spin-spin interactions} (ESSI). 
The Hamiltonian of the ESSI has the form:
\begin{equation}
\mathcal{H}^{*} = \mathcal{H}_{z} + \mathcal{H}^{*}_{d} + \mathcal{H}^{*}_{dd}, 
\label{eqn:2}
\end{equation}
\begin{gather*}
\mathcal{H}_{z} = \hbar\omega_{0}\sum_{j}S^{z}_{j}, ~ \mathcal{H}^{*}_{d} = \hbar A \sum_{fj}S^{z}_{j}S^{z}_{f}, ~
\mathcal{H}^{*}_{dd} = \hbar B \sum_{fj}(S^{-}_{j}S^{+}_{f} + S^{+}_{j}S^{-}_{f}),
\end{gather*}
where $A$ and $B$ defines the average values of parameters 
$\mathrm{A}_{fj}$ and $\mathrm{B}_{fj}$ for separate molecule. 
The Hamiltonian of such type was considered in paper related 
to multiple nonergodic quantum dynamics \cite{bib:7}. It 
seems that the range of applicability of this Hamiltonian 
is much wider \emph{viz.}, it can be applicable for consideration 
the sets of spins in mesoscopic magneto-active clusters. 

The averaged constants $\mathrm{A}$ and $\mathrm{B}$ take into 
account the interaction of the chosen spin with all 
others, keeping the difference between the longitudinal 
($A$) and transverse ($B$) spin components. These constants 
can be obtained from real interaction potential as
\begin{gather*}
\langle \mathrm{A}\rangle =(1/n)\sum_{fj}\mathrm{A}_{fj},  ~
\Delta \mathrm{A} =[(1/n)\sum_{fj}(\mathrm{A}_{fj}-\mathrm{A})^{2}]^{1/2}, ~
\mathrm{A} =\langle \mathrm{A}\rangle\pm \Delta \mathrm{A},	
\end{gather*}
and the same for constant $\mathrm{B}$.

In the next section the energy spectrum of the 
Hamiltonian (\ref{eqn:2}) will be found. 

\section{The matrix structure of the ESSI model}
The operators $\mathcal{H}_{z}$ and $\mathcal{H}^{*}_{d}$ 
commute with each other and with a longitudinal 
component $S^{z}_{j}$ of a separate spin. So the eigenvalues 
\begin{equation}
E_{np} = \hbar\omega_{0}(2p-n)/2 + \hbar A(3p^{2}-3np+n^{2}-n)/4
\label{eqn:3}
\end{equation}
of the operator $\mathcal{H}_{z} +\mathcal{H}^{*}_{d}$ 
are expressed by means of eigenvalues ($\pm 1/2$) of the operator $S^{z}_{j}$, 
where $n$ defines the total number of spins and $p$ is a 
number of spins oriented positively along the applied 
magnetic field. Eigenfunctions of the operator 
$\mathcal{H}_{z} +\mathcal{H}^{*}_{d}$
\begin{multline}
|\chi_{\alpha=1}(n,p)\rangle = \\
= |m_{1}=1/2, m_{2}=1/2,\ldots, m_{p}=1/2, \\
m_{p+1}= -1/2,\ldots, m_{n-1}= -1/2, m_{n}= -1/2\rangle \equiv \\
\equiv |+,+,\ldots,+,-,\ldots,-,-\rangle
\label{eqn:4}
\end{multline}
are also constructed as products of $n$ eigenfunctions 
$|m_{j}\rangle$ of operators $S^{z}_{j}$. The state (\ref{eqn:3}) is degenerated  
$z(n,p) = C^{p}_{n}$ times at the given $n$ and $p$. 
All set of eigenfunctions $|\chi_{\alpha}(n,p)\rangle$, 
belonging to the given $E_{np}$ is obtained from the function  
$|\chi_{\alpha=1}(n,p)\rangle$ by permutation of its 
initial arguments running over all combinations 
from $n$ over $p$, i.e. $\alpha = 1,2,\ldots, z(n,p)$.

Due to the structure of the operators $S^{-}_{j}S^{+}_{f} + S^{+}_{j}S^{-}_{f}$, 
the Hamiltonian $\mathcal{H}^{*}_{dd}$ does not have 
nonzero matrix elements between states with unequal 
number of positively oriented spins. So its matrix 
has blocked-diagonal form, where the sizes of blocks 
is determined as $C^{p}_{n} \times C^{p}_{n}$. Diagonal 
elements of the operator $\mathcal{H}_{z} + \mathcal{H}^{*}_{d}$ 
inside blocks are the same and determined by expression  
(\ref{eqn:3}). In such a way the problem of diagonalization of the 
Hamiltonian (\ref{eqn:2}) is reduced to the finding 
of eigenvalues of the operator $\mathcal{H}^{*}_{dd}$ 
in basis of states (\ref{eqn:4}) with eigenvalues 
belonging to (\ref{eqn:3}).

\section{Eigenvalues of the operator $\mathcal{H}^{*}_{dd}$}
For diagonalization of the operator $\mathcal{H}^{*}_{dd}$ 
a special numerical program was written. It helps to find 
the desired matrices in basis (\ref{eqn:4}) inside 
($n,p$)-blocks and includes subsequent calculation 
of eigenvalues and eigenfunctions of corresponding 
matrices. The spectrum of eigenvalues together with 
their degeneration multiplicity $g$ of the operator 
$\mathcal{H}^{*}_{dd}$ has been found for all $n$ and $p$ 
satisfying to condition $n \le 8$. Analyzing these 
results we established that in each ($n,p$)-block:
\begin{enumerate}
\item the number of different eigenvalues is defined as
\begin{equation}
r(n,p) = r(n,n-p) = p+1,
\label{eqn:5}
\end{equation}
\item the unequal eigenvalues of the operator $\mathcal{H}^{*}_{dd}$ are determined as
\begin{multline}
\hbar\varepsilon_{k}(n,p)=\hbar\varepsilon_{k}(n,n-p) = \hbar B[-p+k(n-2p+1)+k^{2}],
\label{eqn:6}
\end{multline}
\item the energy degeneration of the level $\varepsilon_{k}(n,p)$ is
\begin{equation}
g_{k}(n,p)=g_{k}(n,n-p) = C^{p-k}_{n}-C^{p-k-1}_{n},
\label{eqn:7}
\end{equation}
\end{enumerate}
where $0 \le p \le p^{*}, ~ 0 \le k \le p, ~ p^{*} =[n/2]$ 
is the integer part of $n/2$ and $k$ numbers the operator 
$\mathcal{H}^{*}_{dd}$ unequal eigenvalues.

In further the correctness of these formulae was 
confirmed by direct numerical calculations for 
all possible values of $n$ and $p$ up to $n=11$, $p=4$. 
These calculations give us a ground to consider 
that formulae (\ref{eqn:3}), (\ref{eqn:5}) and (\ref{eqn:6}) 
are exact expressions for the spectrum of the Hamiltonian 
(\ref{eqn:2}) describing ESSI-model.

The eigenfunctions of the Hamiltonian $\mathcal{H}^{*}_{dd}$ 
were also calculated 
\begin{equation}
|\Psi_{\beta}(n,p)\rangle = \sum_{\alpha=1}^{z}C_{\alpha\beta}(n,p)|\chi_{\alpha}(n,p)\rangle, ~ 1 \le \beta \le z(n,p),
\label{eqn:8}
\end{equation}
by determination the sets of coefficients
\begin{displaymath}
(C_{1\beta}(n,p), C_{2\beta}(n,p), C_{3\beta}(n,p),\ldots,C_{z\beta}(n,p)).
\end{displaymath}
Unfortunately, the compact expressions for coefficients 
$C_{\alpha\beta}(n,p)$ were not found. Expressions for 
$C_{\alpha\beta}(n,p)$ were obtained numerically for 
all cases, where the energy spectrum has been determined. 
These expressions are rather cumbersome and so, for example, 
in the Table 1 the coefficients $C_{\alpha\beta}(n,p)$ 
were given only for $n=5$.

\begin{table*}
\label{tab:1}
\caption{The eigenvalues, degeneration multiplicity value and eigenfunctions 
of the Hamiltonian $\mathcal{H}^{*}_{dd}$ for five spins}
\begin{tabular}{|c|c|r|r|l|l|}
\hline
$n$ & $p$ & $\varepsilon_{k}(n,p)$ & $g_{k}(n,p)$ & Coefficients $C_{\alpha\beta}(n,p)$ & Basis functions \\
$ $ & $ $ &                        &              & of the functions $|\Psi_{\beta}(n,p)\rangle$ & $|\chi_{\alpha}(n,p)\rangle$ \\
\hline
5 & 0 & 0 & 1 & (1) & $|-,-,-,-,-\rangle$ \\
\hline
 & 1 & -1 & 4 & $(-1,0,0,0,1)/\sqrt{2}$     & $|+,-,-,-,-\rangle$ \\
 &   &    &   & $(-1,0,0,2,-1)/\sqrt{6}$    & $|-,+,-,-,-\rangle$ \\
 &   &    &   & $(-1,0,3,-1,-1)/2\sqrt{3}$  & $|-,-,+,-,-\rangle$ \\
 &   &    &   & $(-1,4,-1,-1,-1)/2\sqrt{5}$ & $|-,-,-,+,-\rangle$ \\
 &   &  1 & 1 & $(1,1,1,1,1)/\sqrt{5}$      & $|-,-,-,-,+\rangle$ \\
\cline{2-6}
 & 2 & -2 & 5 & $(1,1,-1,-1,-1,0,0,0,0,1)/\sqrt{6}$     & $|+,+,-,-,-\rangle$ \\
 &   &    &   & $(1,-1,1,-1,-1,0,0,0,2,-1)/\sqrt{10}$   & $|+,-,+,-,-\rangle$ \\
 &   &    &   & $(2,-2,-3,3,-2,0,0,5,-1,-2)/2\sqrt{15}$ & $|+,-,-,+,-\rangle$ \\
 &   &    &   & $(-2,2,1,-1,-2,0,4,1,-1,-2)/6$          & $|+,-,-,-,+\rangle$ \\
 &   &    &   & $(-1,1,-1,1,-1,3,-1,-1,1,-1)/3\sqrt{2}$ & $|-,+,+,-,-\rangle$ \\
 &   &  1 & 4 & $(-1,-1,1,1,-2,0,0,0,0,2)/2\sqrt{3}$    & $|-,+,-,+,-\rangle$ \\
 &   &    &   & $(-1,1,-1,1,0,-2,0,0,2,0)/2\sqrt{3}$    & $|-,+,-,-,+\rangle$ \\
 &   &    &   & $(-2,-1,-1,-2,1,1,0,2,1,1)/3\sqrt{2}$   & $|-,-,+,+,-\rangle$ \\
 &   &    &   & $(1,-4,-4,1,1,1,6,-4,1,1)/3\sqrt{10}$   & $|-,-,+,-,+\rangle$ \\
 &   &  6 & 1 & $(1,1,1,1,1,1,1,1,1,1)/\sqrt{10}$       & $|-,-,-,+,+\rangle$ \\
\cline{2-6}
 & 3 & -2 & 5 & $(1,1,-1,-1,0,0,-1,0,0,1)/\sqrt{6}$     & $|+,+,+,-,-\rangle$ \\
 &   &    &   & $(1,-1,-1,1,0,0,-1,0,2,-1)/\sqrt{10}$   & $|+,+,-,+,-\rangle$ \\
 &   &    &   & $(-3,3,-2,2,0,0,-2,5,-1,-2)/2\sqrt{15}$ & $|+,+,-,-,+\rangle$ \\
 &   &    &   & $(1,-1,-2,-2,0,4,2,1,-1,-2)/6$          & $|+,-,+,+,-\rangle$ \\
 &   &    &   & $(-1,1,-1,-1,3,-1,1,-1,1,-1)/3\sqrt{2}$ & $|+,-,+,-,+\rangle$ \\
 &   &  1 & 4 & $(-2,-2,-2,1,1,1,0,0,0,3)/2\sqrt{6}$    & $|+,-,-,+,+\rangle$ \\
 &   &    &   & $(-6,2,2,-5,-5,3,0,0,8,1)/2\sqrt{42}$   & $|-,+,+,+,-\rangle$ \\
 &   &    &   & $(1,-5,2,-5,2,-4,0,7,1,1)/3\sqrt{14}$   & $|-,+,+,-,+\rangle$ \\
 &   &    &   & $(1,1,-4,1,-4,-4,6,1,1,1)/3\sqrt{10}$   & $|-,+,-,+,+\rangle$ \\
 &   &  6 & 1 & $(1,1,1,1,1,1,1,1,1,1)/\sqrt{10}$       & $|-,-,+,+,+\rangle$ \\
\cline{2-6}
 & 4 & -1 & 4 & $(-1,0,0,0,1)/\sqrt{2}$     & $|+,+,+,+,-\rangle$ \\
 &   &    &   & $(-1,0,0,2,-1)/\sqrt{6}$    & $|+,+,+,-,+\rangle$ \\
 &   &    &   & $(-1,0,3,-1,-1)/2\sqrt{3}$  & $|+,+,-,+,+\rangle$ \\
 &   &    &   & $(-1,4,-1,-1,-1)/2\sqrt{5}$ & $|+,-,+,+,+\rangle$ \\
 &   &  1 & 1 & $(1,1,1,1,1)/\sqrt{5}$      & $|-,+,+,+,+\rangle$ \\
\hline
 & 5 & 0 & 1 & (1) & $|+,+,+,+,+\rangle$ \\
\hline
\end{tabular}
\end{table*}

Taking into account the fact that the function 
$|\chi_{\alpha}(n,p)\rangle$ belongs to the same 
eigenvalue (\ref{eqn:3}), one can conclude that any their 
linear combination, in particular combination (\ref{eqn:8}), 
also belongs to this eigenvalue. Due to this property the 
eigenvalues of the ESSI Hamiltonian (\ref{eqn:2}) can be 
presented in the form of simple sum of energies (\ref{eqn:3}) 
and (\ref{eqn:6})
\begin{equation}
\mathcal{E}_{k}(n,p) = E_{np} + \hbar\varepsilon_{k}(n,p),
\label{eqn:9}
\end{equation}
To each set of the three indices $n,p,k$ corresponds
$g_{k}(n,p)$ eigenfunction (\ref{eqn:8}).

In conclusion we would like to note that exact 
expressions (\ref{eqn:3}), (\ref{eqn:5})-(\ref{eqn:7}) 
found for the quantum model ESSI are not exact solution 
but has the \emph{anzatz} status. The found solution was 
guessed and then checked on large number of cases 
including block ($n=11, p=4$), which consists of 330 states.

\section*{}
One of us (A.R.K) expresses his deep acknowledgements 
to Prof. B.M. Fang for an opportunity to know the results 
of experiments of his group before publication.

This work was supported by REC-007 and fond NIOKR RT, 06-6.1-158.


\begin{thebibliography}{7}

\bibitem{bib:1}
A.\,Abragam, The Principles of Nuclear Magnetism, Oxford: Clarendon Press, 1961.

\bibitem{bib:2}
V.\,A. Acarkin, Nuclear dynamic polarization in solid dielectrics, Moscow: Nauka, 1980.

\bibitem{bib:3}
V.\,A. Acarkin, M.\,I. Rodak, Usp. Fiz. Nauk {\bf 107}, 3 (1972).

\bibitem{bib:4}
M. Goldman, Spin temperatire and nuclear magnetic resonance in solids, Oxford: Clarendon Press, 1970.

\bibitem{bib:5}
A.\,K. Khitrin, V.\,L. Ermakov, B.\,M. Fung, Chem. Phys. Lett. {\bf 350}, 161 (2002).

\bibitem{bib:6}
A.\,K. Khitrin, V.\,L. Ermakov, B.\,M. Fung, J. Chem. Phys. {\bf 117}, 6903 (2002).

\bibitem{bib:7}
M.\,G. Rudavec, E.\,B. Fel'dman, Pis'ma Zh. Eksp. Teor. Fiz. {\bf 75}, 760 (2002).

\end{thebibliography}
\end{document}